\pdfoutput=1
\documentclass[sigconf]{acmart}

\AtBeginDocument{%
  }

\providecommand{\correspondingauthor}{}

\copyrightyear{2026}
\acmYear{2026}
\setcopyright{cc}
\setcctype{by}
\acmConference[RecSys '26]{20th ACM Conference on Recommender Systems}{September 27-October 02, 2026}{Minneapolis, MN, USA}
\acmBooktitle{20th ACM Conference on Recommender Systems (RecSys '26), September 27-October 02, 2026, Minneapolis, MN, USA}
\acmDOI{10.1145/3773078.3841249}
\acmISBN{979-8-4007-2284-4/2026/09}

\begin{document}

\title{Should I Be Polite to My LLM Relevance Judge?
  Tone as a Severity Operating-Point Shift}

\author{Tian Zhang}
\correspondingauthor
\orcid{0009-0009-8198-9060} 
\affiliation{%
  \institution{Independent Researcher}
  \city{San Jose}
  \state{California}
  \country{USA}
}
\email{0tianzhang0@gmail.com} 

\author{Meng Li}
\orcid{0009-0008-8866-0201}
\affiliation{%
  \institution{Independent Researcher}
  \city{San Jose}
  \state{California}
  \country{USA}
}
\email{limengunique0917@gmail.com}

\renewcommand{\shortauthors}{Zhang and Li}

\begin{abstract}
Large language models are increasingly used as relevance judges, yet
their labels can shift with prompt surface form. We study one such
feature---tone---on $3{,}498$ TREC DL19/DL20 query--passage pairs,
across eight judge models, five classifier-calibrated politeness levels,
and three paraphrases per level. Effects are strongly
\emph{model-dependent}: one judge shows a structured U-shaped response,
whereas most show only small changes. Where tone changes agreement, the
results are more consistent with a shift in the judge's \emph{severity
operating point}---its overall scoring leniency---than with improved
judgment. Agreement rises or falls as this shift moves the judge toward
or away from human annotators' strictness. A query-disjoint cross-fit retains the expected association (Spearman $\rho=-0.683$; exact model-block permutation $p=0.019$). Tone affects calibration-based agreement more than ranking outcomes: across 32 model--tone contrasts, the largest absolute mean change in NDCG@10 is $0.011$, although Kendall's $\tau$ as low as $0.743$ shows that reordering is reduced, not absent. The account reconciles prior contradictory findings and identifies prompt tone as a potential validity threat when absolute relevance labels matter.
\end{abstract}



\maketitle

\section{Introduction}
\label{sec:intro}
Large language models are now routinely used as relevance judges in
information retrieval and recommendation, assigning graded relevance
labels that stand in for expensive human assessment~\cite{thomas2024searcher, faggioli2023perspectives, upadhyay2024umbrela, zheng2023judging}.
In recommender systems specifically~\cite{lin2025recsys}, LLM judges increasingly serve as
offline evaluators---scoring item relevance, ranking quality, and the
outputs of conversational and generative recommenders---and as a source
of synthetic relevance labels for training and distillation~\cite{shang2025distillation}.
This
makes their reliability consequential---yet a growing body of work
shows that reliability is fragile, shifting with surface-level features
of the prompt such as wording, option order, and
formatting~\cite{arabzadeh2025prompt, sclar2024formatspread}.

One such feature is \emph{tone}: how politely or rudely the prompt
addresses the model. Yet the evidence is both contradictory and
confined to generation tasks: some studies report that rude prompts
\emph{harm} performance~\cite{yin2024respect}, others that they
\emph{help}~\cite{dobariya2025tone}, and a third that the effect is
strongly model-dependent~\cite{cai2025tone}. None studies relevance
judging, and none offers a mechanism explaining \emph{why} tone helps
in one setting and hurts in another. An evaluator today cannot answer a
simple question: should I be polite to my LLM relevance judge?

We answer it and offer an operating-point account. Across eight judges, five classifier-calibrated politeness levels, and paraphrase controls, the effect is strongly model-dependent: one judge shows a large, structured response, whereas most show only small changes. Where tone moves agreement, the pattern is consistent with a shift in the judge's \emph{severity operating point}---the leniency with which it assigns absolute scores---rather than with a uniform gain in discrimination. Tone appears to turn the judge's strictness knob more than its intelligence knob.

Our contribution is threefold. First, we introduce the severity operating-point account, which can explain when tone shifts agreement and reconcile prior contradictory findings: rudeness may help or harm depending on how it moves a model's leniency relative to a reference. Second, we test a pre-specified directional account with query-disjoint cross-fitting: the association persists when alignment and agreement changes are computed on disjoint query folds ($\rho=-0.683$; exact model-block $p=0.019$). Third, we identify a concrete validity threat: tone perturbs absolute, calibration-based metrics such as Cohen's $\kappa$ far more than ranking quality in this setting, though it does not leave rankings invariant.

\section{Method}
\label{sec:method}

\textbf{Task and data.}
Under the UMBRELA protocol~\cite{upadhyay2024umbrela}, each judge assigns
a 0--3 relevance score to a query--passage pair. We use TREC Deep
Learning DL19 and DL20~\cite{craswell2020dl19,craswell2021dl20}, treating
their NIST human relevance judgments (\emph{qrels}) as reference labels.
The frozen dataset contains the $3{,}498$ pairs in the intersection of
BM25 top-50 results and human-annotated pairs. Five cost-efficient models
evaluate all pairs twice; three flagship models use a frozen $40\%$
subsample. One flagship is reasoning-native and reported separately.

\textbf{Calibrated tone with paraphrase control.}
We construct five politeness levels (L1 rude to L5 deferential). The
relevance rubric and output schema are \emph{byte-identical}; only a
wrapper around the fixed rubric varies. Wrappers range from blunt
(L1: ``Score this passage. Don't waste my time with explanations.'') to
highly deferential (L5: ``I would be incredibly grateful if you could
kindly evaluate the relevance of the passage below\ldots''). Intel
\texttt{polite-guard} scores~\cite{intel2024politeguard} are monotonic
across levels ($0.00/0.68/1.00/2.33/3.00$), with no overlap between the
extremes and neutral. Three paraphrases per level ($15$ variants)
separate between-level patterns from wording artifacts.%
\footnote{All 15 prompt wrappers, pre-specified predictions, and
analysis code: \url{https://github.com/dukesky/politeness-llm}.}

\textbf{Operating-point mechanism.}
We measure agreement with linear-weighted Cohen's $\kappa$ between a
judge's scores and the qrels, per level; $\kappa$ reflects how well a
judge's \emph{absolute} score assignments align with humans, and is
thus sensitive to leniency shifts even when the relative ordering is
unchanged. Let $\bar{s}_\ell$ be the
judge's mean score at level $\ell$ and $\bar{s}_{\mathrm q}$ the qrels
mean. We define the \emph{strictness bias}
$\Delta = \bar{s}_{\mathrm{L3}} - \bar{s}_{\mathrm q}$ ($\Delta>0$:
more lenient than humans) and the \emph{tonal drift}
$D(\ell) = \bar{s}_\ell - \bar{s}_{\mathrm{L3}}$. The mechanism predicts
the \emph{direction} of agreement change via the \emph{alignment change}
\begin{equation}
A(\ell) = \bigl|\Delta + D(\ell)\bigr| - \bigl|\Delta\bigr|,
\end{equation}
where $A(\ell)<0$ means the drift moves the judge toward the human
operating point (predicting $\kappa$ to rise) and $A(\ell)>0$ away
(predicting $\kappa$ to fall). The pre-specified prediction is that $A(\ell)$ and $\kappa(\ell)-\kappa(\mathrm{L3})$ are negatively associated. These statistics summarize different aspects of the score matrix---mean severity and pair-level agreement---but are derived from the same underlying observations. We therefore deterministically split queries into two folds, estimate $A(\ell)$ on one fold and the agreement change on the other, swap the fold roles, and combine both directions. Exact inference permutes whole model blocks, preserving the eight cross-fitted contrasts within each model. Kendall's $\tau$ between each tone-level ranking and the L3 ranking provides a complementary check of ranking stability.

\section{Results}
\label{sec:results}

\begin{figure}[t]
\centering
\includegraphics[width=0.92\columnwidth]{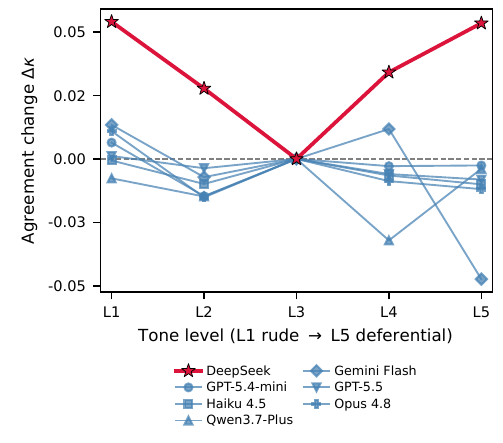}
\caption{Change in agreement $\Delta\kappa$ across all five tone levels.
DeepSeek (red) traces a structured U-shape. Other judges show smaller
or unstructured trajectories; Gemini's L5 drop is driven by one
anomalous paraphrase. Table~\ref{tab:main} reports the two extremes.}
\Description{Line chart of agreement changes from tone levels L1 to L5
for seven non-reasoning models. DeepSeek forms a structured U-shaped
trajectory with positive changes at both extremes. Other trajectories
are smaller or unstructured, including a negative Gemini L5 change.}
\label{fig:dkappa}
\end{figure}

\textbf{Tonal sensitivity is model-dependent.}
Table~\ref{tab:main} reports neutral agreement and changes at the two extremes. DeepSeek is the only judge with a structured effect: lenient at L3 ($\Delta=+0.543$), it becomes stricter at both extremes, raising agreement in a U-shape. Paired query-clustered bootstrap intervals exclude zero for both extreme contrasts: L1 $\Delta\kappa=0.054$, 95\% CI $[0.038,0.070]$, and L5 $\Delta\kappa=0.053$, 95\% CI $[0.040,0.066]$. Its between/within ratio is 1.58. Other changes are small or unstructured, including for Claude Opus despite its ratio above 1 (Figure~\ref{fig:dkappa}). A positive $\Delta\kappa$ alone does not establish improved discrimination.

\begin{figure}[t]
\centering
\includegraphics[width=0.92\columnwidth]{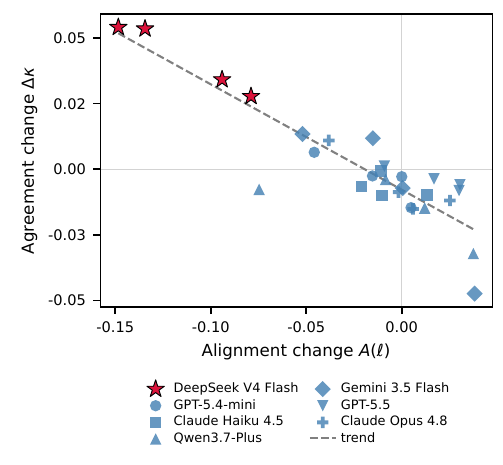}
\caption{Full-data descriptive association across 28 model--tone cells ($\rho=-0.799$). DeepSeek (red) has the largest shifts, but the association remains negative without it ($\rho=-0.680$). Query-disjoint cross-fitting yields $\rho=-0.683$ with exact model-block $p=0.0187$ (Table~\ref{tab:spearman}).
}
\Description{Scatter plot of alignment change against agreement change
across 28 model--tone cells. The dashed trend line slopes downward,
and the DeepSeek points show the largest positive agreement changes.}
\label{fig:spearman}
\end{figure}

\textbf{The observed association is consistent with the pre-specified account.}
All seven non-reasoning judges are lenient at baseline ($\Delta>0$), consistent with documented LLM-judge leniency~\cite{arabzadeh2025benchmarking}. In the full 28-cell data, $A(\ell)$ and the agreement change have a descriptive Spearman correlation of $\rho=-0.799$. Under query-disjoint cross-fitting, the two fold directions give $\rho=-0.684$ and $-0.685$; combined, $\rho=-0.683$, with exact model-block permutation $p=0.019$ (Figure~\ref{fig:spearman}; Table~\ref{tab:spearman}). The full-data association remains negative after excluding the anomalous Gemini L5 cell ($\rho=-0.775$), DeepSeek ($\rho=-0.680$), or both ($\rho=-0.636$).

\textbf{Tone perturbs calibration far more than ranking.}
Across 32 model--tone contrasts, the largest absolute mean $\Delta\mathrm{NDCG}@10$ was $0.011$, and only one unadjusted query-bootstrap 95\% interval excluded zero. Mean Kendall's $\tau$ against L3 ranged from $0.743$ to $0.947$; the lowest estimate had a 95\% CI of $[0.722,0.764]$. Ranking quality therefore changed far less than $\kappa$, although reordering was not absent.

\textbf{Paraphrase controls separate level-wide from idiosyncratic sensitivity.}
One DeepSeek L2 paraphrase received a lower classifier politeness score ($0.534$ versus $0.73$--$0.77$), behaved more strictly, and raised $\kappa$ as predicted. Conversely, one Gemini 3.5 Flash L5 paraphrase, despite receiving the same classifier politeness score as its siblings ($2.999$), alone reduced agreement ($\kappa:0.45\rightarrow0.27$). We treat this as an idiosyncratic wording sensitivity rather than evidence of a level-wide tone effect.

\section{Discussion and Conclusion}
\label{sec:discussion}
A higher $\kappa$ need not mean better judgment. DeepSeek is lenient at
L3, while both tone extremes make it stricter and raise agreement. This
is consistent with movement toward the stricter qrels---an alignment
that could reverse against another reference.

The account can reconcile reports that rude prompts harm
performance~\cite{yin2024respect} or help it~\cite{dobariya2025tone}:
the direction depends on the model and reference operating points. For
RecSys pipelines, tone poses a larger threat when LLM judges certify
absolute labels or generate training targets. Ranking outcomes changed
far less in our setting, but were not immune. Where absolute values matter, calibrate tone and vary surface form;
when ranking is the target, report rank-stability checks rather than
assuming invariance.

On the reasoning-native judge, rude prompts used $120\%$ more reasoning
tokens than neutral yet produced the lowest agreement, consistent with
evidence that more reasoning need not improve passage
reranking~\cite{jedidi2025overthink}.

Our claims have four limitations: the operating-point account remains
observational despite query-disjoint validation; only one politeness
classifier, seven non-reasoning models, and three paraphrases per level
limit generalization; no equivalence margin establishes ranking safety;
and the rude wrappers also alter instructions about explanation effort,
so tone is not perfectly isolated from instruction content. The account
therefore describes a robust association, not an independently identified
mechanism or a general guarantee of ranking safety.

\begin{table}[t]
\centering
\caption{Per-model sensitivity. $\Delta\kappa$ compares each extreme
with neutral L3; ``btw/in'' is between/within-level variation.
Only DeepSeek shows a structured U-shape; Gemini~3.1~Pro is
reasoning-native. \textsuperscript{*}One anomalous Gemini L5$_a$
paraphrase; see Section~\ref{sec:results}.}
\label{tab:main}
\small
\begin{tabular}{@{}lrrrr@{}}
\toprule
Model & $\kappa_{\mathrm{L3}}$ & $\Delta\kappa_{\mathrm{L1}}$ & $\Delta\kappa_{\mathrm{L5}}$ & btw/in \\
\midrule
DeepSeek V4 Flash      & 0.430 & $+0.054$ & $+0.053$ & 1.58 \\
GPT-5.4-mini           & 0.327 & $+0.006$ & $-0.003$ & 0.71 \\
Claude Haiku 4.5       & 0.434 & $-0.001$ & $-0.010$ & 0.30 \\
Qwen3.7-Plus           & 0.486 & $-0.008$ & $-0.004$ & 0.80 \\
Gemini 3.5 Flash       & 0.441 & $+0.013$ & $-0.047^{*}$ & 0.93 \\
GPT-5.5                & 0.374 & $+0.001$ & $-0.008$ & 0.73 \\
Claude Opus 4.8        & 0.497 & $+0.011$ & $-0.012$ & 1.37 \\
\midrule
Gemini 3.1 Pro         & 0.445 & $-0.023$ & $-0.006$ & 2.35 \\
\bottomrule
\end{tabular}
\end{table}

\begin{table}[t]
\centering
\caption{Association between alignment change $A(\ell)$ and
agreement change $\kappa(\ell)-\kappa(\mathrm{L3})$. Full-data
results are descriptive. Cross-fitting estimates the two quantities
on disjoint query folds; the reported exact $p$-value permutes whole
model blocks, preserving eight cross-fitted contrasts per model.}
\label{tab:spearman}
\small
\begin{tabular}{@{}lrrr@{}}
\toprule
Analysis & $n$ & $\rho$ & exact $p$ \\
\midrule
Full data                    & 28 & $-0.799$ & -- \\
Fold 0 $\rightarrow$ Fold 1  & 28 & $-0.684$ & -- \\
Fold 1 $\rightarrow$ Fold 0  & 28 & $-0.685$ & -- \\
Combined cross-fit           & 56 & $-0.683$ & $0.0187$ \\
\bottomrule
\end{tabular}
\end{table}

\bibliographystyle{ACM-Reference-Format}
\bibliography{references}


\begin{thebibliography}{16}


\ifx \showCODEN    \undefined \def \showCODEN     #1{\unskip}     \fi
\ifx \showISBNx    \undefined \def \showISBNx     #1{\unskip}     \fi
\ifx \showISBNxiii \undefined \def \showISBNxiii  #1{\unskip}     \fi
\ifx \showISSN     \undefined \def \showISSN      #1{\unskip}     \fi
\ifx \showLCCN     \undefined \def \showLCCN      #1{\unskip}     \fi
\ifx \shownote     \undefined \def \shownote      #1{#1}          \fi
\ifx \showarticletitle \undefined \def \showarticletitle #1{#1}   \fi
\ifx \showURL      \undefined \def \showURL       {\relax}        \fi
\providecommand\bibfield[2]{#2}
\providecommand\bibinfo[2]{#2}
\providecommand\natexlab[1]{#1}
\providecommand\showeprint[2][]{arXiv:#2}

\bibitem[Arabzadeh and Clarke(2025)]%
        {arabzadeh2025prompt}
\bibfield{author}{\bibinfo{person}{Negar Arabzadeh} {and} \bibinfo{person}{Charles L.~A. Clarke}.} \bibinfo{year}{2025}\natexlab{}.
\newblock \showarticletitle{A Human-{AI} Comparative Analysis of Prompt Sensitivity in {LLM}-Based Relevance Judgment}. In \bibinfo{booktitle}{\emph{Proceedings of the 48th International ACM SIGIR Conference on Research and Development in Information Retrieval}}.
\newblock
\showeprint[arxiv]{2504.12408}


\bibitem[Arabzadeh et~al\mbox{.}(2025)]%
        {arabzadeh2025benchmarking}
\bibfield{author}{\bibinfo{person}{Negar Arabzadeh}, \bibinfo{person}{Ehsan Kamalloo}, \bibinfo{person}{Xinyu Zhang}, {and} \bibinfo{person}{Charles L.~A. Clarke}.} \bibinfo{year}{2025}\natexlab{}.
\newblock \bibinfo{title}{Benchmarking {LLM}-based Relevance Judgment Methods}.
\newblock
\showeprint[arxiv]{2504.12558}~[cs.IR]


\bibitem[Cai et~al\mbox{.}(2025)]%
        {cai2025tone}
\bibfield{author}{\bibinfo{person}{Jiawei Cai} {et~al\mbox{.}}} \bibinfo{year}{2025}\natexlab{}.
\newblock \bibinfo{title}{Does Tone Change the Answer?}
\newblock
\showeprint[arxiv]{2512.12812}~[cs.CL]


\bibitem[Craswell et~al\mbox{.}(2021)]%
        {craswell2021dl20}
\bibfield{author}{\bibinfo{person}{Nick Craswell}, \bibinfo{person}{Bhaskar Mitra}, \bibinfo{person}{Emine Yilmaz}, {and} \bibinfo{person}{Daniel Campos}.} \bibinfo{year}{2021}\natexlab{}.
\newblock \showarticletitle{Overview of the {TREC} 2020 Deep Learning Track}. In \bibinfo{booktitle}{\emph{Proceedings of the Twenty-Ninth Text REtrieval Conference (TREC 2020)}}.
\newblock
\showeprint[arxiv]{2102.07662}


\bibitem[Craswell et~al\mbox{.}(2020)]%
        {craswell2020dl19}
\bibfield{author}{\bibinfo{person}{Nick Craswell}, \bibinfo{person}{Bhaskar Mitra}, \bibinfo{person}{Emine Yilmaz}, \bibinfo{person}{Daniel Campos}, {and} \bibinfo{person}{Ellen~M. Voorhees}.} \bibinfo{year}{2020}\natexlab{}.
\newblock \showarticletitle{Overview of the {TREC} 2019 Deep Learning Track}. In \bibinfo{booktitle}{\emph{Proceedings of the Twenty-Eighth Text REtrieval Conference (TREC 2019)}}.
\newblock
\showeprint[arxiv]{2003.07820}


\bibitem[Dobariya and Kumar(2025)]%
        {dobariya2025tone}
\bibfield{author}{\bibinfo{person}{Kirtan Dobariya} {and} \bibinfo{person}{Abhishek Kumar}.} \bibinfo{year}{2025}\natexlab{}.
\newblock \bibinfo{title}{Mind Your Tone}.
\newblock
\showeprint[arxiv]{2510.04950}~[cs.CL]


\bibitem[Faggioli et~al\mbox{.}(2023)]%
        {faggioli2023perspectives}
\bibfield{author}{\bibinfo{person}{Guglielmo Faggioli}, \bibinfo{person}{Laura Dietz}, \bibinfo{person}{Charles L.~A. Clarke}, \bibinfo{person}{Gianluca Demartini}, \bibinfo{person}{Matthias Hagen}, \bibinfo{person}{Claudia Hauff}, \bibinfo{person}{Norbert Fuhr}, \bibinfo{person}{Benno Stein}, \bibinfo{person}{Johanne~R. Trippas}, {and} \bibinfo{person}{Nicola Ferro}.} \bibinfo{year}{2023}\natexlab{}.
\newblock \showarticletitle{Perspectives on Large Language Models for Relevance Judgment}. In \bibinfo{booktitle}{\emph{Proceedings of the 2023 ACM SIGIR International Conference on Theory of Information Retrieval (ICTIR)}}.
\newblock
\showeprint[arxiv]{2304.09161}


\bibitem[{Intel}(2024)]%
        {intel2024politeguard}
\bibfield{author}{\bibinfo{person}{{Intel}}.} \bibinfo{year}{2024}\natexlab{}.
\newblock \bibinfo{title}{Polite Guard}.
\newblock \bibinfo{howpublished}{\url{https://huggingface.co/Intel/polite-guard}}.
\newblock


\bibitem[Jedidi et~al\mbox{.}(2025)]%
        {jedidi2025overthink}
\bibfield{author}{\bibinfo{person}{Nour Jedidi}, \bibinfo{person}{Yung-Sung Chuang}, \bibinfo{person}{James Glass}, {and} \bibinfo{person}{Jimmy Lin}.} \bibinfo{year}{2025}\natexlab{}.
\newblock \bibinfo{title}{Don't ``Overthink'' Passage Reranking: Is Reasoning Truly Necessary?}
\newblock
\showeprint[arxiv]{2505.16886}~[cs.IR]


\bibitem[Lin et~al\mbox{.}(2025)]%
        {lin2025recsys}
\bibfield{author}{\bibinfo{person}{Jianghao Lin}, \bibinfo{person}{Xinyi Dai}, \bibinfo{person}{Yunjia Xi}, \bibinfo{person}{Weiwen Liu}, \bibinfo{person}{Bo Chen}, \bibinfo{person}{Hao Zhang}, \bibinfo{person}{Yong Liu}, \bibinfo{person}{Chuhan Wu}, \bibinfo{person}{Xiangyang Li}, \bibinfo{person}{Chenxu Zhu}, \bibinfo{person}{Huifeng Guo}, \bibinfo{person}{Yong Yu}, \bibinfo{person}{Ruiming Tang}, {and} \bibinfo{person}{Weinan Zhang}.} \bibinfo{year}{2025}\natexlab{}.
\newblock \showarticletitle{How Can Recommender Systems Benefit from Large Language Models: A Survey}.
\newblock \bibinfo{journal}{\emph{ACM Transactions on Information Systems}} \bibinfo{volume}{43}, \bibinfo{number}{2} (\bibinfo{year}{2025}), \bibinfo{pages}{1--47}.
\newblock


\bibitem[Sclar et~al\mbox{.}(2024)]%
        {sclar2024formatspread}
\bibfield{author}{\bibinfo{person}{Melanie Sclar}, \bibinfo{person}{Yejin Choi}, \bibinfo{person}{Yulia Tsvetkov}, {and} \bibinfo{person}{Alane Suhr}.} \bibinfo{year}{2024}\natexlab{}.
\newblock \showarticletitle{Quantifying Language Models' Sensitivity to Spurious Features in Prompt Design or: How I Learned to Start Worrying about Prompt Formatting}. In \bibinfo{booktitle}{\emph{The Twelfth International Conference on Learning Representations (ICLR)}}.
\newblock
\showeprint[arxiv]{2310.11324}


\bibitem[Shang et~al\mbox{.}(2025)]%
        {shang2025distillation}
\bibfield{author}{\bibinfo{person}{Hongwei Shang}, \bibinfo{person}{Nguyen Vo}, \bibinfo{person}{Nitin Yadav}, \bibinfo{person}{Tian Zhang}, \bibinfo{person}{Ajit Puthenputhussery}, \bibinfo{person}{Xunfan Cai}, \bibinfo{person}{Shuyi Chen}, \bibinfo{person}{Prijith Chandran}, {and} \bibinfo{person}{Changsung Kang}.} \bibinfo{year}{2025}\natexlab{}.
\newblock \showarticletitle{Knowledge Distillation for Enhancing {Walmart} E-commerce Search Relevance Using Large Language Models}. In \bibinfo{booktitle}{\emph{Companion Proceedings of the ACM Web Conference 2025}}. \bibinfo{pages}{449--457}.
\newblock


\bibitem[Thomas et~al\mbox{.}(2024)]%
        {thomas2024searcher}
\bibfield{author}{\bibinfo{person}{Paul Thomas}, \bibinfo{person}{Seth Spielman}, \bibinfo{person}{Nick Craswell}, {and} \bibinfo{person}{Bhaskar Mitra}.} \bibinfo{year}{2024}\natexlab{}.
\newblock \showarticletitle{Large Language Models can Accurately Predict Searcher Preferences}. In \bibinfo{booktitle}{\emph{Proceedings of the 47th International ACM SIGIR Conference on Research and Development in Information Retrieval}}.
\newblock
\showeprint[arxiv]{2309.10621}


\bibitem[Upadhyay et~al\mbox{.}(2024)]%
        {upadhyay2024umbrela}
\bibfield{author}{\bibinfo{person}{Shivani Upadhyay}, \bibinfo{person}{Ronak Pradeep}, \bibinfo{person}{Nandan Thakur}, \bibinfo{person}{Daniel Campos}, \bibinfo{person}{Nick Craswell}, \bibinfo{person}{Ian Soboroff}, \bibinfo{person}{Hoa~Trang Dang}, {and} \bibinfo{person}{Jimmy Lin}.} \bibinfo{year}{2024}\natexlab{}.
\newblock \bibinfo{title}{{UMBRELA}: {UMbrela} is the {Bing} {RELevance} Assessor}.
\newblock
\showeprint[arxiv]{2406.06519}~[cs.IR]


\bibitem[Yin et~al\mbox{.}(2024)]%
        {yin2024respect}
\bibfield{author}{\bibinfo{person}{Ziqi Yin}, \bibinfo{person}{Hao Wang}, \bibinfo{person}{Kaito Sugimoto}, \bibinfo{person}{Chenhan Yuan}, \bibinfo{person}{Nguyen~Ngoc Duc}, \bibinfo{person}{Fei Cheng}, {and} \bibinfo{person}{Sadao Kurohashi}.} \bibinfo{year}{2024}\natexlab{}.
\newblock \showarticletitle{Should We Respect {LLMs}? {A} Cross-Lingual Study on the Influence of Prompt Politeness on {LLM} Performance}. In \bibinfo{booktitle}{\emph{Proceedings of the 1st Workshop on Social Influence in Conversations (SICon) at ACL 2024}}. \bibinfo{pages}{2--14}.
\newblock
\showeprint[arxiv]{2402.14531}


\bibitem[Zheng et~al\mbox{.}(2023)]%
        {zheng2023judging}
\bibfield{author}{\bibinfo{person}{Lianmin Zheng}, \bibinfo{person}{Wei-Lin Chiang}, \bibinfo{person}{Ying Sheng}, \bibinfo{person}{Siyuan Zhuang}, \bibinfo{person}{Zhanghao Wu}, \bibinfo{person}{Yonghao Zhuang}, \bibinfo{person}{Zi Lin}, \bibinfo{person}{Zhuohan Li}, \bibinfo{person}{Dacheng Li}, \bibinfo{person}{Eric~P. Xing}, \bibinfo{person}{Hao Zhang}, \bibinfo{person}{Joseph~E. Gonzalez}, {and} \bibinfo{person}{Ion Stoica}.} \bibinfo{year}{2023}\natexlab{}.
\newblock \showarticletitle{Judging {LLM}-as-a-Judge with {MT-Bench} and {Chatbot} {Arena}}. In \bibinfo{booktitle}{\emph{Advances in Neural Information Processing Systems (NeurIPS) Datasets and Benchmarks Track}}.
\newblock
\showeprint[arxiv]{2306.05685}


\end{thebibliography}

\end{document}